\documentstyle[preprint,eqsecnum,aps]{revtex}
\begin{document}
\draft
\title{Pinning and Tribology of Tethered Monolayers on Disordered
Substrates}
\date{\today}
\author{Carlo Carraro$^a$ and David R. Nelson$^b$}
\address{${}^a$ Department of Chemistry, University of California, 
Berkeley, CA 94720}
\address{${}^b$ Department of Physics, Harvard University, 
Cambridge, MA 02138} 
\maketitle
\begin{abstract}
We study the statistical mechanics and dynamics of crystalline
films with a fixed internal connectivity on a random substrate. 
Defect free triangular lattices exhibit a sharp transition to a
low temperature glassy phase with anomalous phonon fluctuations and a
nonlinear force-displacement law with a continuously variable exponent,
similar to the vortex glass phase of directed lines in 1+1 dimensions. 
The periodicity of the tethered monolayer acts like a filter which
amplifies particular Fourier components of the disorder. However, the absence
of annealed topological defects like dislocations is crucial: 
the transition is destroyed when the constraint of fixed 
connectivity is relaxed and 
dislocations are allowed to proliferate. 
\end{abstract}
\pacs{46.30Pa 68.35Rh 76.60Ge}

\section{Introduction}

The pinning of elastic media subjected to external forces 
is a subject
of considerable interest in connection with a variety of phenomena, ranging 
from epitaxial surface growth to transport in planar arrays of flux lines 
in type II superconductors\cite{[1]}. 
High temperature superconductors are especially
interesting in this respect because of the simultaneous presence of large
thermal fluctuations and of quenched disorder. Much effort has been devoted 
to the study of 1+1 dimensional models, which are models of vortex lines 
confined to a plane\cite{[2],[3]} (see Fig.~1a). 
Although some quantitative questions have yet 
to be answered satisfactorily, a clear qualitative 
picture of the physics involved has emerged, which can be summarized as
follows: the 1+1 dimensional flux array, subjected to external 
point disorder, displays a transition, at some temperature $T_g$, 
between a high temperature regime, dominated by thermal fluctuations, 
and a low temperature regime, where
the behavior of the system is controlled by a line of fixed points.
This is a transition to a disorder-dominated phase, where the
elastic system is pinned. Pinning affects both static correlations and dynamic
responses in a nontrivial way, giving rise to nonlinear
current-voltage characteristics. 
Crucial to pinning is the discrete nature of the elastic system, 
which, roughly speaking, acts as a Fourier filter for components of 
disorder on length scales corresponding to the lattice spacing
(i.e., the distance between flux lines).

The question naturally arises, whether coherent amplification of the disorder 
(assumed to exist at all physically relevant wave lengths) 
is instrumental to the pinning of a vortex array; that is, 
it is important to question the role of long range crystalline order of the
array in selecting out particular Fourier components of the pinning potential.
{\it This feature is built into the} 1+1 {\it dimensional vortex line model}, 
which possesses {\it algebraic} crystalline order at all nonzero 
temperatures {\it and} is also 
topologically perfect. The topological perfection arises for vortex lines 
because the average magnetic field is 
parallel to the plane, and thus the lines cannot terminate in the plane, 
and their labelling in a perfect crystalline sequence is always unambiguous. 

Alternative two dimensional 
models of elastic solids can be considered, which
allow for topological defects, such as dislocations. These are 
2+0 dimensional models (see Fig.~1b). 
For vortices in superconductors, the average 
magnetic field must then be perpendicular to the plane of the film, 
and defects leading to multivalued displacements of the vortices are allowed, 
as they are in many other experimental situations. Additional
experimental realizations include   
colloidal crystals\cite{[4]}, amphiphillic mono- or bi-layers composed of
lipid molecules
(possibly polymerized)\cite{[5]},  electrons in semiconductor
heterostructures\cite{[6]}, 
and magnetic bubble arrays\cite{[7]}. The constraint of fixed nearest neighbor
connectivity could be enforced in some cases by polymerization or, more
generally, simply by large kinetic barriers to particle exchanges at low
temperature.
These systems can be topologically perfect, or else can be subject 
to either quenched or annealed internal defects, 
in addition to the external pinning potential. 

The goal of the present work is to develop the theory of pinned two 
dimensional crystals within the framework of two dimensional elasticity 
theory. 
For the case of a topologically perfect crystal of identical tethered
particles, subjected to thermal
fluctuations and quenched pinning, the behavior is qualitatively similar to
the 1+1 dimensional model (displaying, e.g., a sharp
phase transition to a low temperature pinned phase). The possibility of 
topological defects, however, introduces significant modifications. 

Annealed dislocations destroy positional quasi long 
range order in two dimensional crystals above some finite melting temperature 
$T_{M}$\cite{[8]}, where a liquid crystalline ``hexatic'' phase exists,
with algebraic long range order in the bond angle. 
As will be shown below, the melting temperature is always smaller than
$T_g$, so that a two dimensional crystal is always melted at the 
temperature below which pinning disorder would become relevant for a
topologically perfect solid (i.e., dislocations are a relevant 
perturbation at $T_g$). Thus, the phase transition discussed above 
is washed out in the presence of thermally excited dislocation pairs.
Analogous behavior is
encountered in the random field XY model\cite{[3]}. An extension to vector
displacement fields has been studied by Giamarchi and LeDoussal\cite{[9]}.  

The hexatic liquid crystalline phase of the {\it untethered} membrane above 
$T_M$ (see Fig.~2a) 
displays similar behavior when subjected to a component of the random
substrate disorder which couples directly 
to the bond angle field. The analogy with the random field XY
model\cite{[3]} becomes a rigorous mapping for annealed hexatic membranes:
either disclination unbinding or substrate disorder always distabilize the
hexatic line of fixed points, and it is unclear if there is a sharp
finite temperature phase transition. 
Polymerized tethered membranes behave differently, however.
Although quenched-in unpaired dislocations destroy translational
long range order, they cannot drive the shear modulus to zero. The finite
shear modulus makes the bond angle fluctuations ``massive''\cite{[8]}. These
fluctuations are now stable to weak external disorder. 

{\it Quenched} topological disorder has recently been studied for ``tethered
surfaces''\cite{[10]}, in which defects are frozen into a 
two-dimensional network
of covalently bonded particles fluctuating in three dimensions. It is of
considerable interest to determine what happens when such disordered 
tethered surfaces are forced to lie flat and brought into contact 
with a disordered polycrystalline or amorphous substrate. A particularly
simple example of such tethered disorder is shown in Fig.~2b, where a
topologically perfect triangular lattice is disrupted by random
substitutional disorder. Unfortunately, the method used in this paper cannot
be directly applied to such systems\cite{[thanks]}. Tethered substitutional
disorder invalidates a straightforward analogy with random field models (see
Sec.~2). Cule and Hwa have studied this problem in one dimension, and
concluded that a new, stronly pinned glassy phase arises, characterized by
exponents in the ``random manifold'' universality class\cite{[rm]}. Similar
results may apply to two dimensional tethered networks with quenched-in
vacancies, interstitials, dislocations or disclinations as well as random
substitutional disorder\cite{[rm]}.

A theory of {\it three} dimensional tethered networks with both quenched
random internal defects {\it and} a quenched random external potential would
have interesting implications for the tangled arrays of vortex lines which
may arise when bulk Type II superconductors are
subjected to strong external magnetic fields. If melted flux liquids are
cooled rapidly, barriers to flux cutting\cite{[11]} may become sufficiently large  
that the vortex lines freeze at low temperatures into a nonequilibrium
directed ``polymer glass''\cite{[12]}. The usual triangular Abrikosov
flux lattice would then be disrupted by a quenched array of dislocation and
disclination lines of arbitrarily large size. The resulting vortex array
could have a shear modulus over a wide range of experimental time scales
because of entanglement constraints, and
would be subject to point pinning by imperfections in the underlying host
superconducting material. 

This paper is organized as follows. In Sec.~II, we present the theory of 
ideal (topologically perfect) two dimensional crystals subject to external point 
disorder. Effects of annealed dislocations are discussed in Sec.~III, and
some concluding remarks are presented in Sec.~IV. 
Some of the effects of quenched substitutional disorder are
discussed in Appendix A. Technical details of the derivation of the
renormalization group recursion relations used in Sec.~II are contained
in Appendix B. Appendix C derives the exponent $z$ for a relaxational model
of tethered crystalline membrane dynamics. 
An Ornstein-Zernicke description  of the hexatic phase is 
derived in Appendix D.

\section{Pinning of ideal crystals}

Consider a two dimensional Bravais lattice 
$\vec R_{mn}=m\vec a+n\vec b$ 
and denote by $\vec u(m,n)$ the displacement from the equilibrium 
position $\vec R_{mn}$. We begin by restricting our analysis to an ideal 
lattice of identical particles, 
where the displacements are single-valued functions of position. 
Thus, we exclude for the moment the possibility of topological defects.
The strains associated with random substitutional disorder will be discussed
later in this section. 
The energy of an ideal crystal undergoing a small deformation can be
expressed, using continuum elasticity theory, by the harmonic hamiltonian 
\begin{equation}
H_0={1\over 2}\int d^2r (\lambda u_{ii}^2 +2\mu u_{ij}^2),
\label{eq:1}
\end{equation}
where $\lambda$ and $\mu$ are Lam\'e coefficients and 
the strain tensor is defined as
\begin{equation}
u_{ij}(\vec r)={1\over 2}(\partial_iu_j+\partial_ju_i).
\label{eq:2}
\end{equation}
The only regular lattice in two dimensions with sufficient symmetry to be
described by the {\it isotropic} elastic theory (\ref{eq:1}) is a triangular array
\cite{[16]}. In this case, it is convenient to take 
\begin{equation}
\vec a=a_0(1,0),\ \ \vec b=a_0\biggl({1\over 2},{\sqrt{3}\over 2}\biggr),
\end{equation}
where $a_0$ is the lattice constant.

If the lattice is subjected to an external pinning potential, 
the total pinning
energy is given by $U_{pin}=\sum_{mn}V[\vec R_{mn}+\vec u(m,n)]$. 
With the aid of Poisson's summation formula, we write 
\begin{eqnarray}
U_{pin}&&=\int d^2rV(r)\sum_{mn}\delta^2[\vec r-\vec R_{mn}-\vec u(m,n)]
\nonumber\\ &&=\int d^2rV(r)\int d\sigma_1\int d\sigma_2 
\delta^2[\vec r-\sigma_1\vec a-\sigma_2\vec b-\vec u(\sigma_1, \sigma_2)]
\sum_{pq}e^{2\pi i(p\sigma_1+q\sigma_2)},
\label{eq:3}
\end{eqnarray}
where $(m,n)$ and $(p,q)$ are pairs of integers.
Next, we change integration variables from $\sigma_1,\sigma_2$ 
to $\vec r=(x,y)=\sigma_1\vec a+\sigma_2\vec b+\vec u(\sigma_1, \sigma_2)$. 
The Jacobian of this tranformation is 
\begin{equation}
{\partial \sigma_1\over\partial x}{\partial \sigma_2\over\partial y}-
{\partial \sigma_1\over\partial y}{\partial \sigma_2\over\partial x}
\simeq{1\over |\vec a\times \vec b|}(1-\vec\nabla\cdot\vec u).
\end{equation}
The $\delta$ function in Eq.~(\ref{eq:3}) fixes the value of $\sigma$ to be the root of
\begin{equation}
\sigma_1\vec a+\sigma_2\vec b+\vec u(\sigma_1, \sigma_2)=\vec r,
\label{eq:4a}
\end{equation}
which can be expanded in series of the small displacement $\vec u$.
Upon noting that the zeroth order term is 
\begin{equation}
\sigma_1^0={(\vec r \times \vec b)\cdot\hat {\bf z}\over 
|\vec a \times \vec b|}\ ;
\ \ \ \ 
\sigma_2^0=-{(\vec r \times \vec a)\cdot\hat {\bf z}
\over |\vec a \times \vec b|}\ ,
\label{eq:4b}
\end{equation}
and using the $\delta$ function to eliminate the integrals over $\sigma_1$ and
$\sigma_2$, we can write the total pinning energy as 
\begin{equation}
U_{pin}\approx\int d^2rV(r){1\over |\vec a\times \vec b|} 
(1-\vec\nabla\cdot\vec u)
\sum_{mn}e^{i\vec G_{mn}\cdot[\vec r-\vec u(\vec r)]},
\label{eq:5}
\end{equation}
where the $\{\vec G_{mn}\}$ are reciprocal lattice vectors. The total 
hamiltonian of the system is
obtained by adding the hamiltonian of an ideal two
dimensional crystal to the pinning energy.
We study the effect of a random distribution of weak 
pinning potentials, with mean and variance defined by
\begin{equation}
\overline{V(\vec r)}=0, \ \ \ \ 
\overline{V(\vec r)V(\vec r^\prime)}=\Delta\delta^2(\vec r-\vec r^\prime).
\label{eq:6}
\end{equation}

If we restrict our attention to $\vec G_{mn}=0$ and to the six smallest
reciprocal lattice vectors in the summation of Eq.~(5), then the total
Hamiltonian which comprises our model may be approximated (up to an additive
constant) by
\begin{equation}
{\cal H}={1\over 2}\int d^2r \biggl(\lambda u_{ii}^2 +2\mu u_{ij}^2-w(\vec r)u_{ii}+
\sum_{l=1}^3V_{G_l}(\vec r)e^{-i\vec G_{l}\cdot\vec u(\vec r)}
\biggr),
\label{eq:7}
\end{equation}
where the $\vec G_{l}$, $l=1,2,3,$ are three reciprocal lattice vectors
inclined at 
$120^{\rm o}$ angles to each other in the innermost ring,
\begin{equation}
w(\vec r)={V(\vec r)\over |\vec a\times\vec b|},
\end{equation}
and $V_{\vec G}(\vec r)$ is a local Fourier component of the random
potential
\begin{equation}
V_{\vec G}(\vec r)={1\over \Delta \Omega}\int d^2r'
e^{i\vec G\cdot\vec r'}{V(\vec r')\over |\vec a\times\vec b|}.
\label{eq:8}
\end{equation}
The integration is carried out over an area $\Delta \Omega$, centered on
$\vec r$, large compared to the lattice spacing but small compared to the
sample dimensions. We have neglected terms of the form $\vec\nabla \cdot\vec
u\exp(i\vec G_{l}\cdot[\vec r-\vec u(\vec r)])$, which are less important
than the ones we have kept. 
In Appendix A, we show that random substitutional disorder contributes to
$w(\vec r)$ and leads as well to a random term of the form $-{1\over
2}\int d^2r w_{ij}(\vec r)u_{ij}(\vec r)$. A two dimensional bead and spring
model with random spring lengths contains similar contributions. Quenched
substitutional disorder has important additional effects, however. Internal
disorder in the particle sizes or bond lengths violates the {\it discrete}
translational invariance of the Hamiltonian (2.10) under $\vec u(\vec r)\to
\vec u(\vec r)+\vec R_{mn}$. Cule and Hwa have argued that this
symmetry-breaking leads to an effective ``random manifold'' potential which
depends in a complicated way on {\it both} $\vec r$ and $\vec u(\vec r)$, and
causes an instability to a more strongly pinned glassy phase than the one
studied here \cite{[rm]}.

For tethered networks of {\it identical} particles, 
we expect that pinning effects due to the random phases and amplitudes
embodied in the $\{V_{G_l}\}$ will be important below some critical 
temperature, and that the properties of the system in the vicinity 
of this temperature will be perturbatively accessible
by renormalization group methods.

The starting point of the renormalization program is perturbation theory.
Computation of the disorder-averaged
observables of interest, such as the free energy or the two-point functions,
requires expanding the logarithm of the partition function $\cal Z$ in powers
of the weak pinning potential and averaging term by term. 
Such averages are conveniently handled by the replica trick, which involves
first calculating ${\cal Z}^n=1+n\log {\cal Z} + O(n^2)$ and eventually taking
the limit $n\to 0$.
The disorder average of ${\cal Z}^n$ leads to 
\begin{equation}
\overline{{\cal Z}^n }= 
\int\!\! {\cal D}\vec u_1 \dots {\cal D}\vec u_n\ 
e^{-{{\cal H}^0_n\over T}-{{\cal H}^I_n\over T}},
\label{eq:9}
\end{equation}
where the harmonic term is
\begin{equation}
{{\cal H}^0_n\over T}=
{1\over 2T}\sum_{\alpha \beta}^n\int{d^2k\over (2\pi)^2} 
u_{i\alpha}(-\vec k)k^2\biggl(\biggl(\mu P^T_{ij}+
(2\mu+\lambda)P^L_{ij}
\biggr)\delta_{\alpha \beta} -BP^T_{ij}-AP^L_{ij}\biggr)
u_{j\beta}(\vec k),
\label{eq:10}
\end{equation}
and the interaction term is
\begin{equation}
{{\cal H}^I_n\over T}=-g\sum_{l=1}^3\int {d^2r}\sum_{\alpha \beta}^n
\cos\bigl[\vec G_l\cdot (\vec u_\alpha(\vec r)-\vec u_\beta(\vec r))\bigr].
\label{eq:11}
\end{equation}
Greek indices label different replicas, while latin subscripts 
($i,j=1,2$) are used for the components of the displacement vector $\vec u$. 
Summation over repeated indices $i,j$ is understood.
The vectors $\vec G_l$ are the smallest nonzero vectors in the reciprocal
lattice. For a triangular lattice of spacing $a_0$, $l=1,2,3$ 
and $|G_l|^2=16\pi^2/3a_0^2$. Effects due to reciprocal lattice 
vectors of larger norm are irrelevant.

The structure of the replicated Hamiltonian is quite simple. The harmonic part
contains a term diagonal in the replica indices. This is simply the replicated
hamiltonian of an ideal two dimensional crystal in Fourier space, where 
$P^L_{ij}$ and $P^T_{ij}$ are
longitudinal and transverse projectors, respectively. 
In addition, transverse and longitudinal terms are
considered, which are constant in replica space. The tranverse term, while not
present initially, is generated by renormalization. Its coefficient $B$ can be
set to zero initially. 
The cosine term arises as a consequence of the discrete nature of the lattice.
Its amplitude is related to the correlation function of the randomness by
\begin{equation}
g={\Delta\over T^2\Omega^2},
\label{eq:12}
\end{equation}
where $\Omega=|\vec a\times \vec b|$ is the area of the unit cell.

The use of replicas brings out an important property of the model.
Note that the interaction term involves only differences of fields with
different replica indices. Hence, the ``center of mass,'' in replica space, 
of the fields $\vec u_\alpha$, 
\begin{equation}
\vec \Psi \equiv n^{-1/2}\sum_\alpha \vec u_\alpha,
\label{eq:13}
\end{equation} 
is a free field and does not suffer renormalization. This symmetry implies 
that $2\mu+\lambda-nA,\ \mu-nB$ do not renormalize, a result valid to all 
orders in perturbation theory, similar to the invariance under 
renormalization  of the spin wave stiffness in the random field XY model
\cite{[17]}. As a consequence, the renormalization group 
flow of the disorder coupling, $g$, is one dimensional. 

Next, consider the connected Green's function 
\begin{equation}
\overline{<u_i(\vec r)u_j(\vec r^\prime)>_c}
=-{\partial^2 \overline{\log{\cal Z}} \over 
\partial J_i(\vec r) \partial J_j(\vec r^\prime)}
=-{1\over n}{\partial^2 \overline{{\cal Z}^n[\vec J]} 
\over \partial J_i(\vec r) \partial J_j(\vec r^\prime)}\biggr|_{n=0},
\label{eq:14}
\end{equation}
obtained from the generating functional 
\begin{equation}
\overline{{\cal Z}^n[\vec J]}
=\overline{{\cal Z}^n}<\exp\biggl(i\int d^2r
\vec J(\vec r) \cdot\sum_\alpha \vec u_\alpha (\vec r)\biggr)>_R.
\label{eq:15}
\end{equation}
Hereafter, the notation $<\ >_R$ will stand for average with respect 
to the integrand in Eq.~(\ref{eq:9}). 
Since the source $\vec J$ couples only to $\vec\Psi$, a free field, 
this Green's function is the free correlation function, {\rm independent of
g},
\begin{equation}
\overline{<u_i(\vec r)u_j(\vec r^\prime)>_c}=-\delta_{ij}{T\over 4\pi \mu}
{3\mu+\lambda\over 2\mu+\lambda } 
\log{|\vec r-\vec r^\prime|\over a_0} + {\rm const},\ \ \ 
|\vec r -\vec r_0|\to\infty,
\label{eq:16}
\end{equation}
where the connected part is $<A(x)B(x')>_c=<A(x)B(x')>-<A(x)><B(x')>$.

The peculiar properties of the glassy phase do appear, however, in the
nontrivial behavior of some response functions as well as in the full
correlation function 
$\overline{<u_i(\vec r)u_j(\vec r^\prime)>}$, which, according to the
analysis above, is a probe of sample-to-sample fluctuations. In the language
of replicas, these fluctuations are captured by introducing a replica
dependent source field, $\vec J_\alpha(\vec r)$, and a new
generating functional
\begin{equation}
\overline{{\cal Z}^n[\{\vec J_\alpha\}]} =\overline{{\cal Z}^n}
<\exp\biggr(i\int d^2r
\sum_\alpha \vec J_\alpha(\vec r) \cdot\vec u_\alpha (\vec r)\biggl)>_R.
\label{eq:17}
\end{equation}
Differentiation with respect to $\vec J_\alpha$ yields the Green's functions
\begin{equation}
G_{ij\alpha \beta}(\vec r -\vec r^\prime)= 
<u_{i\alpha}(\vec r)u_{j\beta}(\vec r^\prime)>_R.
\label{eq:18}
\end{equation}
Provided symmetry under permutation of the replica indices holds, we can write
\begin{eqnarray}
\overline{<u_i(\vec r)u_j(\vec r^\prime)>_c}&&=
\lim_{n\to 0}[G_{ij11}(\vec r-\vec r^\prime)-G_{ij12}(\vec r-\vec r^\prime)]
\nonumber\\
\overline{<u_i(\vec r)u_j(\vec r^\prime)>}&&=
\lim_{n\to 0}G_{ij11}(\vec r-\vec r^\prime),
\label{eq:19}
\end{eqnarray}
where the limit $n\to 0$ is here simply a convenient bookkeeping device 
for doing perturbation theory.

The perturbation series for the disorder-averaged Green's
functions diverges in the thermodynamic limit at low temperature.
The divergent diagrams are most easily recognized 
by considering the expansion of the free energy,
\begin{equation}
\overline{(F-F_0)}=
-\lim_{n\to 0}{1\over n}\biggl(-<{\cal H}^I_n>_R+
{1\over 2}\bigl(<({\cal H}^I_n)^2>_R
-<({\cal H}^I_n)>^2_R\bigr)+\dots\biggr).
\label{eq:20}
\end{equation}
Upon defining a reduced temperature 
\begin{equation}
\tau\equiv 1-{T\over T_g}
=1-{T|G|^2\over 8\pi\mu}{3\mu+\lambda\over 2\mu+\lambda},
\label{eq:21}
\end{equation}
one finds, up to regular terms, in order $g$: 
\begin{equation}
-{<{\cal H}^I_n>_R\over T}\sim gca^2n(n-1)3\pi(L/a\sqrt{c})^{2\tau},
\label{eq:22}
\end{equation}
and in order $(g^2)$:
\begin{equation}
{1\over 2T^2}<({\cal H}^I_n)^2>_R \sim 
-g^2c^2a^4n(n-1)[I_0(\psi)+(n-2)I_0(\psi/2)] 3\pi^2
\biggl({L\over a\sqrt{c}}\biggr)^{2\tau}
{2\over 2\tau}\biggl(\bigl(L/a\sqrt{c}\bigr)^{2\tau}-1\biggr).
\label{eq:23}
\end{equation}
Infrared and ultraviolet cutoffs $L$ and $a$, respectively, have been
introduced; $c= {1\over 4}e^{2E}\approx 0.79$, where $E$ is Euler's constant; 
$I_0$ is a modified Bessel function, and $\psi\equiv 
{T|G|^2(\mu+\lambda)\over 4\pi\mu(2\mu+\lambda)}$.
The details of the calculation can be found in Appendix B. 
 
The divergences can be removed order by order in a double expansion in 
powers of $g$ and $\tau$. The parameters of
the renormalized theory transform, under rescaling of length by $e^l$,
according to the following equations:
\begin{eqnarray}
{d\lambda\over dl}&&=0\nonumber\\
{d\mu\over dl}&&=0\nonumber\\
{d \tilde g\over dl}&&=2\tau \tilde g-{2\over 3}{\tilde g}^2
[2I_0(\psi/2)-I_0(\psi)]\nonumber\\
{dA\over dl}&&
={\tilde g}^2 {4\over 3}\mu{2\mu+\lambda\over 3\mu +\lambda}
\bigl[I_0(\psi)-{1\over 2}I_1(\psi)\bigr]\nonumber\\
{dB\over dl}&&
={\tilde g}^2 {4\over 3}\mu{2\mu+\lambda\over 3\mu +\lambda}
\bigl[I_0(\psi)+{1\over 2}I_1(\psi)\bigr],
\label{eq:24}
\end{eqnarray}
where $\tilde g \equiv 3\pi gca^2(L/a\sqrt{c})^{2\tau}$.
The flow of the disorder coupling to zero for $\tau<0$, i.e., for $T>T_g$, 
means that the discreteness of the lattice is irrelevant in the high 
temperature phase. Thus, the correlation functions in this phase are 
similar to the gaussian model of Ref.\cite{[18]}. In particular,
\begin{equation}
\overline{<u_i(\vec r)u_i(0)>}\sim -\eta\log r
\label{eq:25}
\end{equation}
where
\begin{equation}
\eta=|\vec G|^2\biggl({T\over 4\pi \mu}{3\mu+\lambda\over 2\mu+\lambda } 
+{\Delta\over 4\pi(2\mu+\lambda)^2}+{A\over 4\pi(2\mu+\lambda)^2} +
{B\over 4\pi\mu^2} \biggr).
\label{eq:26}
\end{equation}

Below $T_g$, on the other hand, the disorder coupling flows toward a finite
fixed point value of $\tilde g^*=3\tau/[2I_0(\psi/2)-I_0(\psi)]$. 
The runaway flows of $A$ and $B$
cause the correlation function to grow as  $\log^2r$, a behavior which 
was termed ``superroughening'' in the context of scalar models of 
surface growth\cite{[19]}.

As we anticipated, another distinctive property of the glassy phase is the
nontrivial near-equilibrium dynamics. The dissipative dynamics 
of the system embodied in the Langevin equation
\begin{equation}
\gamma\partial_t\vec u=-{\delta H\over\delta\vec u}+\vec\zeta,
\label{eq:27}
\end{equation}
with $\vec\zeta$ a thermal noise, can also be studied by dynamical 
renormalization group methods 
\cite{[20]}. The detailed treatment of this model is described in
Appendix C. Regularization 
of the perturbative expansion of the dynamic response leads to 
a renormalized friction coefficient $\gamma$, from which the dynamic 
exponent can be extracted
\begin{eqnarray}
z&&=2,\ \ \ \ \ \ \ \ \ \ \ \ T>T_g\nonumber\\
z&&=2+{24\over\sqrt{c}}{2\mu+\lambda\over 3\mu+\lambda}
\biggl({\mu\over 2\mu+\lambda}\biggr)^{\mu\over 3\mu+\lambda}
{\tau\over [2I_0(\psi/2)-I_0(\psi)]},\ \ \ T<T_g,
\label{eq:28}
\end{eqnarray}
where $c={1\over 4}e^{2E}\approx 0.79$ is the same function of Euler's
constant as appears in the static calculation.

\section{EXTERNAL DISORDER AND TOPOLOGICAL DEFECTS}
The foregoing discussion assumed the fixed 
topology of an ideal lattice. 
Spontaneous nucleation of topological defects, which occurs above the
melting temperature
\begin{equation}
T_M={\mu\over 4\pi}{\mu+\lambda\over 2\mu+\lambda}a^2,
\label{eq:29}
\end{equation}
destroys translational long range order\cite{[8]}.
The ratio of the glass temperature to the melting temperature is always 
greater than one (in fact, $T_g/T_M\ge 6$ for all elastic constant values in
the physically relevant range 
$\mu > 0,\  \mu+\lambda > 0$), so that dislocations are expected to be a 
strongly relevant perturbation at $T_g$. Thus, {\it either} the random
substrate potential or thermally excited dislocation pairs always
destabilize the harmonic hamiltonian, and the transition of the previous
section does not occur in the presence of annealed topological defects. 

Although annealed dislocations destroy translational long range order, the
resulting hexatic phase does possess (algebraic) long range order in the bond
angle. A harmonic hamiltonian for the hexatic phase can be obtained in
Ornstein-Zernicke approximation, valid at long wavelength. The details of the
derivation are presented in Appendix D. We can consider the stability of the
long wave length hamiltonian, Eq.~(D14), to an external random potential
coupled to the bond angle. An experimental realization of this system is
provided by a hexatic liquid crystalline film adsorbed onto a polycrystalline
substrate, that is, a substrate whose randomly varying crystallographic axes
locally bias the orientation of the bonds in the film. The total hamiltonian
becomes precisely that of a random field XY model, which was studied by Cardy
and Ostlund \cite{[3]}. Similar to the case of {\it untethered} crystalline
films discussed above, it follows from ref. \cite{[3]} that the harmonic
hamiltonian (D14) is always destabilized either by the external disorder or
by thermally excited {\it disclinations}. There is an important
difference between crystalline and hexatic membranes, however. In the case of a
crystalline membrane, the ideal topology can be fixed by polymerization, and
the fixed line discovered in Sec.~II should be experimentally accessible. In
contrast, it is impossible to prevent disclinations from destabilizing the
vortex glass fixed line in a hexatic film, except by quenching the
topology of the film. But in this case, Eq.~(D14) reveals, upon treating the
singular density and bond angle fluctuations, $\delta\rho_s$ and 
$\delta\theta_s$, as quenched variables, that this quenched hexatic phase has
a finite shear modulus, rendering the bond fluctuations massive. 
Because of the finite shear modulus, the bond angle excitations 
are stable to weak disorder. Whether a tethered hexatic (or liquid) film is
unstable to a ``random manifold'' glassy state \cite{[rm]} is an interesting
subject for future investigation.

\section{DISCUSSION AND CONCLUSIONS}

In this paper, we have studied the physics of 2+0 dimensional
arrays of identical particles 
in an external random potential, using a vector extension 
of the 1+1 dimensional random phase model. The predictions for 
vector models without topological defects are qualitatively similar 
to the scalar model: the discreteness of the vortex
array coherently enhances the Fourier components of the external disorder
which are commensurate with the lattice, leading to a glassy phase at low 
temperature. 
This phase is characterized by a nonlinear response to an 
external driving force, as well as by static correlations which diverge 
more strongly than a simple logarithm. Quantitative expressions for the 
static and dynamic exponents near the transition were computed within a 
perturbative renormalization group scheme. 

The physics changes completely, if annealed topological defects are allowed
in the two dimensional lattice. The possibility of important topological
defects constitutes the principal difference
between the 2+0 and the 1+1 dimensional models of vortex arrays. 
Both annealed and quenched dislocations have been considered, 
the most interesting case being provided by quenched dislocations. 
The nonvanishing shear modulus of a membrane with quenched-in
dislocations prevent the bond angle from following the random bias of an
external polycrystalline substrate, so that the bond angle order
parameter in quenched hexatics is stable to weak disorder of this type.

Our study is relevant to several other situations besides the pinning of 
vortex arrays in type-II superconductors. Systems of current experimental
interest were mentioned in the Introduction. Here, we would like to comment
on possible applications to tribology, the study of friction and
lubrication. We are interested in the behavior of two surfaces brought 
in contact and rubbed against each other in the presence of an intermediate
thin layer of lubricant. This boundary layer is often modelled as a two
dimensional, incommensurate crystalline overlayer\cite{[22]}. 
Our work may be useful in generating 
more realistic descriptions which allow for a) surface imperfections,
acting as pinning centers on the lubricant overlayer; b) changes in 
topology of the overlayer, especially excitation of dislocations, which must
surely be important at finite temperature and/or under finite stresses.
We leave the pursuit of this interesting topic to future work.

{\it Noted added}:  After this paper was submitted, we received an interesting
preprint by D. Carpentier and P. Le Doussal (cond-mat/9611168) which reaches
similar conclusions  using a different renormalization group method. 
Comparison with their results enabled us to uncover an error in the first
version of our paper which, although it did not affect our basic conclusions,
changes the coefficients in our recursion relations.   Once the error is
corrected, results obtained by the different methods agree.  We are grateful
to P. Le Doussal for bringing this discrepancy to our attention.

We acknowledge helpful discussions with T. Hwa.
This work was supported by the National Science Foundation, at Harvard
University principally through the Harvard Materials Research and
Engineering Center through Grant No.~DMR91-06237 and through Grant 
No.~DMR91-15491, and at UC Berkeley through Grant No.~CHE 9508336. 
Support by the Office of Naval Research under Grant
N00014-92-J-1361 at UC Berkeley is also gratefully acknowledged.

\appendix
\section*{A}
Effects due to a disordered {\it substrate} were incorporated into
isotropic two dimensional elasticity theory in Section~II. In this Appendix,
we discuss effects on the elastic properties due to random substitutional disorder {\it in the membrane},
as exemplified by the large impurity atom displayed in Fig.~2b. 
We do {\it not} discuss the important interplay between random substitutional
disorder and the disorder substrate potential\cite{[rm]}. A
distribution of impurity atoms with sizes different from the average leads to
random strains. {\it Annealed} defects of this kind can be integrated out and
simply alter the elastic constants $\mu$ and $\lambda$. As we shall see, the
strains in the quenched case contribute to the coefficients $A$ and $B$
displayed in the replicated hamiltonian, Eq.~(\ref{eq:10}). Quenched random vacancy or
interstitial defects as well as tightly bound dislocations pairs or triplets
would affect $A$ and $B$ similarly. We work with a continuum model studied
already in the context of random tethered surfaces fluctuating in three
dimensions\cite{[10]}. The only change required is neglect of displacements normal
to the average plane of the membrane. These phonon modes become massive due
to the interaction with the substrate and can be integrated out without
affecting our basic results. 

We assume a topolgically perfect lattice and replace $\vec R_{mn}$ by a
coarse-grained function $\vec R(\vec r)$ which gives the lattice
displacement  $\vec R$ as a function of the reference position $\vec r$. We
use a generalization of Eq.~(\ref{eq:1}), 
\begin{equation}
H={1\over 2}\int d^2r (\lambda u_{ii}^2 +2\mu u_{ij}^2),
\label{eq:A.1}
\end{equation}
where the strain matrix is now given by\cite{[10]}
\begin{equation}
u_{ij}={1\over 2}\bigl(\partial_i\vec R\cdot\partial_j\vec R
-\partial_i\vec R^0\cdot\partial_j\vec R^0\bigr).
\label{eq:A.2}
\end{equation}
Here, $\vec R^0(\vec r)$ is a preferred lattice distortion which minimizes
the energy in the absence of thermal fluctuations. In the absence of defects,
$\partial_i\vec R^0\cdot\partial_j\vec R^0=\delta_{ij}$. Localized defects
like substitutional disorder, vacancies, interstitials, etc., lead to
deviations which we parameterize by 
\begin{equation}
\partial_i\vec R^0\cdot\partial_j\vec R^0=\delta_{ij}+c_{ij}(\vec r).
\label{eq:A.3}
\end{equation}
If we assume uncorrelated gaussian disorder, the probability distribution of
the tensor $c_{ij}(\vec r)$ takes the form\cite{[10]}
\begin{equation}
{\cal P}_r[c_{ij}(\vec r)]\propto\exp\biggl(-{1\over 2\sigma_1}
\int d^2rc_{ii}^2-{1\over 2\sigma_2}\int d^2rc_{ij}^2\biggr).
\label{eq:A.4}
\end{equation}

We now set $\vec R(\vec r)=\vec r+\vec u(\vec r)$, so that
\begin{equation}
\partial_i\vec R\cdot\partial_j\vec R=\delta_{ij}+
{1\over 2}(\partial_iu_j+\partial_ju_i)+
{1\over 2}\partial_i\vec u\cdot\partial_j\vec u\approx \delta_{ij}+u_{ij}.
\label{eq:A.5}
\end{equation}
The hamiltonian (\ref{eq:A.1}) then takes the form
\begin{equation}
H=const. +{1\over 2}\int d^2r \bigl(\lambda u_{ii}^2 +2\mu u_{ij}^2
-\lambda c_{ii}(\vec r)u_{jj}(\vec r)-2\mu c_{ij}(\vec r)u_{ij}\bigr).
\label{eq:A.6}
\end{equation}
The term proportional to $u_{ii}$ represents random dilations or contractions
due to isolated impurities in positions of high symmetry, while the more
complicated tensorial coupling describes more anisotropic defect
configurations. Upon replicating this hamiltonian and tracing out the
Gaussian disorder, we obtain contributions to the coefficients $A$ and $B$ in
Eq.~(\ref{eq:10}).

\section*{B} 
This Appendix details the calculations leading to the flow equations
for $g$, $A$, and $B$. We begin by evaluating the right hand side of 
Eq.~(\ref{eq:20}) term by term. To order $g$, 
\begin{eqnarray}
\biggl<{{\cal H}^I_n\over T}\biggr>&&=
-g\sum_{l=1}^3\int {d^2r}\sum_{\alpha \beta}^n \biggl<\exp\biggl(i
\vec G_l\cdot (\vec u_\alpha(\vec r)-\vec u_\beta(\vec r))\biggr)\biggr>\nonumber\\
&&=-g\sum_{l=1}^3\int {d^2r}\sum_{\alpha \beta}^n
\exp\biggl(-T\int {d^2k\over (2\pi)^2}{1\over k^2}
G^l_i\bigl({1\over \mu}P^T_{ij}+
{1\over 2\mu +\lambda}P^L_{ij}\bigr)G^l_j\biggr)
\label{eq:B.1}
\end{eqnarray}
Before proceeding with the calculation, we must introduce cutoffs 
to deal with infrared and ultraviolet singularities arising from integrals
of the type
\begin{equation}
\int {d^2k\over (2\pi)^2}G^l_i{e^{i\vec k\cdot \vec r}
\over k^2}P^{T,L}_{ij}G^l_j.
\label{eq:B.2}
\end{equation}
A long wave length cutoff $L$ is introduced 
to eliminate infrared divergences. Its effect amounts to shifting 
$1/k^2\to 1/(k^2+L^{-2})$. The limit $L\to \infty$ can be taken safely 
at the end of the calculations. The ultraviolet divergences are removed in
coordinate space by the simple shift $r\to \sqrt{r^2+a^2}$, where $a$ is a
short wave length cutoff of order the lattice constant. 
Note the short distance limit:
\begin{eqnarray}
\int {d^2k\over (2\pi)^2}
{e^{i\vec k\cdot\vec s}\over k^2+L^{-2}}\biggr|_{s^2=r^2+a^2}&&=
{1\over 2\pi}K_0\bigl({\sqrt{r^2+a^2}\over L})\nonumber\\
&&\to -{1\over 4\pi}\log c{(r^2+a^2)\over L^2}, \ \ \ \ {r\over L}\ll 1.
\label{eq:B.3}
\end{eqnarray}
Here, $K_0$ is a modified Bessel function, and $c=(1/4)e^{2\gamma}$, where
$\gamma$ is Euler's constant. Thus, the asymptotic behavior 
of the integrals in Eq.~(\ref{eq:B.2}) is readily evaluated
\begin{equation}
\int {d^2k\over (2\pi)^2}G^l_i{e^{i\vec k\cdot \vec r}
\over k^2}P^{T,L}_{ij}G^l_j \to -{|G|^2\over 8\pi}\log{c(r^2+a^2)\over L^2}
\pm\biggl({(\vec G\cdot \vec r)^2\over 4\pi r^2}-{|G|^2\over 8\pi}\biggr).
\label{eq:B.4}
\end{equation}
Upon substituting into Eq.~(\ref{eq:B.1}), and recalling Eq.~(\ref{eq:21}), 
which defines the reduced temperature $\tau\equiv 1-{T\over T_g}
=1-{T|G|^2\over 8\pi\mu}{3\mu+\lambda\over 2\mu+\lambda}$, 
we obtain Eq.~(\ref{eq:22}).
 
Next, in order $g^2$, one has
\begin{eqnarray}
\biggl<\biggl({{\cal H}^I_n\over T}\biggr)^2\biggr>&&=
g^2\sum_{l=1}^3\sum_{l'=1}^3\sum_{\alpha \beta}^n\sum_{\alpha' \beta'}^n 
\int\! {d^2r}\!\int\! {d^2r'}\biggl<\exp\biggl(i
\vec G_l\cdot (\vec u_\alpha(\vec r)-\vec u_\beta(\vec r))\biggr)\nonumber\\
&&\phantom{=g^2\sum_{l=1}^3\sum_{l'=1}^3
\int \!{d^2r}\!\int\! {d^2r'}\sum_{\alpha \beta}^n\sum_{\alpha' \beta'}^n}
\times\exp\biggl(i\vec G_{l'}\cdot 
(\vec u_\alpha'(\vec r')-\vec u_\beta'(\vec r'))\biggr)\biggr>\nonumber\\
&&=g^2\sum_{l=1}^3\sum_{l'=1}^3\sum_{\alpha \beta}^n
\sum_{\alpha' \beta'}^n \biggl({\tilde a\over L}\biggr)^{(4-4\tau)}\nonumber\\
&&\phantom{=g^2\sum_{l=1}^3}\!\!\times
\int\!{d^2r}\!\int\!{d^2r'}
\exp\biggl(\!-TQ_{\alpha\beta}^{\alpha'\beta'}
{\vec G_l\cdot \vec G_{l'}\over 8\pi\mu}{3\mu+\lambda\over 2\mu+\lambda}
\log{c(|\vec r-\vec r'|^2+a^2)\over L^2}\biggr)\nonumber\\
&&\phantom{=g^2\sum_{l=1}^3\int\!{d^2r}}
\times\exp\biggl[\!-TQ_{\alpha\beta}^{\alpha'\beta'}
{(\mu+\lambda)\over 4\pi\mu (2\mu+\lambda)}
\biggl(\vec G_l\cdot (\hat r-\hat r')
\vec G_{l'}\cdot (\hat r-\hat r')-{1\over 2}\biggr)\biggr],
\label{eq:B.5}
\end{eqnarray}
where $Q_{\alpha\beta}^{\alpha'\beta'}=\delta_{\alpha\alpha'}+
\delta_{\beta\beta'}-\delta_{\alpha\beta'}-\delta_{\alpha'\beta}$, and use
has been made of Eq.~(\ref{eq:B.3}) in anticipation of the thermodynamic limit.
Hereafter, $\tilde a\equiv\sqrt{c}a$ and 
$\psi\equiv{T|G|^2(\mu+\lambda)\over 4\pi\mu(2\mu+\lambda)}$.

Logarithmic divergencies appear, at the critical temperature ($\tau=0$), 
when $\vec G_l\cdot \vec G_{l'}
Q_{\alpha\beta}^{\alpha'\beta'}/|G|^2=1$. 
The logarithmically divergent terms contribute to the multiplicative
renormalization of the coupling $g$, while all other terms can be absorbed
into an additive constant. (Had we chosen to renormalize the self energy, the
need of an additive renormalization would not have arisen, but the
calculations would have been considerably more involved.)
To proceed, we first fix $l$, $\alpha$, and $\beta$, which can be done in
$3n(n-1)$  independent ways. Then, logarithmic contributions arise either 
for $l'=l$, $Q_{\alpha\beta}^{\alpha'\beta'}=1$, or for $l'\ne l$, 
$Q_{\alpha\beta}^{\alpha'\beta'}=-2$. The respective
combinatorial weights are $2(n-2)$ and $2$. Carrying out the remaining
integrals leads to Eq.~(\ref{eq:23}).

We now compute the flow of the coupling constants $A$, $B$ 
under renormalization group transormations. Because the self energy is
momentum independent in the first order of perturbation theory, the lowest
order contribution to the perturbative renormalization of $A$, $B$ arises in
$O(g^2)$.
The free propagator in momentum space, obtained from Eq.~(\ref{eq:10}), 
can be written as the sum of two terms:
\begin{eqnarray}
G^0_{ij\alpha\beta}(k)&&=M_{ij}(k)\delta_{\alpha\beta}+N_{ij}(k)\nonumber\\
M_{ij}(k)&&={1\over k^2}\biggl({1\over\mu}P^T_{ij}+
{1\over 2\mu+\lambda}P^L_{ij}\biggr)\nonumber\\
N_{ij}(k)&&={1\over k^2}{B\over\mu(\mu-nB)}P^T_{ij}+
{1\over k^2}{A\over (2\mu+\lambda)(2\mu+\lambda-nA)}P^L_{ij}.
\label{eq:B.6}
\end{eqnarray}
The perturbative expansion of the generating functional Eq.~(\ref{eq:17}) through
second order reads as
\begin{eqnarray}
\overline{{\cal Z}^n[\{\vec J_\alpha\}]}&&=
\exp\biggl(-{T\over 2}\int{d^2k\over (2\pi)^2}\sum_\alpha\sum_{ij}
J_{i\alpha}(\vec k) M_{ij}(k)J_{j\alpha}(-\vec k)+
\sum_{\alpha\beta}J_{i\alpha}(\vec k) N_{ij}(k)J_{j\beta}(-\vec k)\biggr)\nonumber\\
&&\times \biggl(1+{g}{\cal F}_1(J)+{g^2\over 2}{\cal
F}_2(J)+\dots\biggr)
\label{eq:B.7}
\end{eqnarray}
where
\begin{eqnarray}
{\cal F}_2(J)&&=\sum_{ll'}\sum_{\alpha\beta}\sum_{\alpha'\beta'}
\int d^2r\int d^2r'\biggl({\tilde a\over L}\biggr)^{(4-4\tau)}
\exp\biggl(\!-TQ_{\alpha\beta}^{\alpha'\beta'}
\sum_{ij}G^l_iM_{ij}(\vec r-\vec r') G^{l'}_j\biggr)\nonumber\\
&&\times\exp\biggl[-T\int{d^2k\over (2\pi)^2}\biggl(
\sum_{ij}G^l_iM_{ij}(k)\bigl(J_{j\alpha}(\vec k)-J_{j\beta}(\vec k)\bigr)
e^{-i\vec k\cdot\vec r}+\nonumber\\
&&\phantom{\times\exp\biggl[-T\int{d^2k\over (2\pi)^2}\biggl(}+
\sum_{ij}G^{l'}_iM_{ij}(k)\bigl(J_{j\alpha'}(\vec k)-J_{j\beta'}(\vec k)\bigr)
e^{-i\vec k\cdot\vec r'}\biggr)\biggr].
\label{eq:B.8}
\end{eqnarray}
Thus, the renormalized propagator $N^R_{ij}(k)$ can be calculated
perturbatively as
\begin{equation}
-TN^R_{ij}(k)\delta(\vec k+\vec k')=-TN_{ij}(k)\delta(\vec k+\vec k')+
{(2\pi)^2g^2\over 2}{\delta^2{\cal F}_2(J)
\over\delta J_{i\lambda}(\vec k)\delta J_{j\mu}(\vec k')}\biggr|_{J=0}+\cdots,
\label{eq:B.9}
\end{equation}
with
\begin{eqnarray}
{\delta^2{\cal F}_2(J)
\over\delta J_{i\lambda}(\vec k)\delta J_{j\mu}(\vec k')}\biggr|_{0}\!
&&=\sum_{ll'}\!\sum_{\alpha\beta}\!\sum_{\alpha'\beta'}\!
\int\!\! d^2r\!\!\int\!\! d^2r'\biggl({\tilde a\over L}\biggr)^{(4-4\tau)}\!\!\!
\exp\biggl(\!-TQ_{\alpha\beta}^{\alpha'\!\beta'}\!
\sum_{i'j'}G^l_{\!i'}M_{i'\!j'}(\vec r-\vec r') G^{l'}_{\!j'}\biggr)\nonumber\\
&&\times\!{T^2\over (2\pi)^4}\!\!\sum_{i'j'}\!\biggl(\!
G^l_{\!i'}M_{i'i}(k)(\delta_{\alpha\lambda}\!-\!\delta_{\beta\lambda})
e^{-i\vec k\cdot\vec r}\!+\!
G^{l'}_{\!i'}M_{i'i}(k)(\delta_{\alpha'\!\lambda}\!-\!\delta_{\beta'\!\lambda})
e^{-i\vec k\cdot\vec r'}\!\biggr)\nonumber\\
&&\times\!\biggl(\
G^l_{j'}M_{j'j}(k')(\delta_{\alpha\mu}-\delta_{\beta\mu})
e^{-i\vec k'\cdot\vec r}+
G^{l'}_{j'}M_{j'j}(k')(\delta_{\alpha'\!\mu}-\delta_{\beta'\!\mu})
e^{-i\vec k'\cdot\vec r'}\!\biggr)
\label{eq:B.10}
\end{eqnarray}
Contributions behaving as $1/k^2$ for small $k$ arise from 
\begin{eqnarray}
&&\sum_{ll'}\!\sum_{\alpha\beta}\!\sum_{\alpha'\beta'}\!\int\!\! 
d^2r\!\!\int\!\! d^2r'\biggl({\tilde a\over L}\biggr)^{(4-4\tau)}\!\!\!
\exp\biggl(\!-TQ_{\alpha\beta}^{\alpha'\!\beta'}\!
\sum_{ij}G^l_iM_{ij}(\vec r-\vec r') G^{l'}_j\biggr)\nonumber\\
&&\times {T^2\over (2\pi)^4}\sum_{mm'}\biggl(
G^l_mM_{mi}(k)(\delta_{\alpha\lambda}-\delta_{\beta\lambda})
e^{-i\vec k\cdot\vec r}
G^{l'}_{m'}M_{m'j}(k')(\delta_{\alpha'\mu}-\delta_{\beta'\mu})
e^{-i\vec k'\cdot\vec r'}+\nonumber\\
&&\phantom{\times {T^2\over (2\pi)^4}\sum_{mm'}\biggl(} 
G^{l'}_mM_{mi}(k)(\delta_{\alpha'\lambda}-\delta_{\beta'\lambda})
e^{-i\vec k\cdot\vec r'}
G^l_{m'}M_{m'j}(k')(\delta_{\alpha\mu}-\delta_{\beta\mu})
e^{-i\vec k'\cdot\vec r}\biggr)\nonumber\\
&&=\delta(\vec k+\vec k')
\sum_{ll'}\!\sum_{\alpha\beta}\!\sum_{\alpha'\beta'}\!
\int\! d^2s \biggl({\tilde a\over L}\biggr)^{(4-4\tau)}\!\!
\exp\biggl(\!-TQ_{\alpha\beta}^{\alpha'\!\beta'}\!
\sum_{ij}G^l_iM_{ij}(s) G^{l'}_j\biggr)\nonumber\\
&&\times{-T^2\over (2\pi)^2}\bigl(\vec k\cdot\vec s\bigr)^2
\sum_{mm'}G^l_mG^{l'}_{m'}M_{mi}M_{m'j}
(\delta_{\alpha\lambda}-\delta_{\beta\lambda})
(\delta_{\alpha'\mu}-\delta_{\beta'\mu})
\label{eq:B.11}
\end{eqnarray}

Power counting reveals that logarithmic divergencies arise when $l=l'$ 
and $\alpha=\beta'$ and $\beta=\alpha'$. Thus, with the help of the following
relations, 
\begin{eqnarray}
\sum_lG^l_mG^l_{m'}={3\over 2}|G|^2\delta_{mm'}\nonumber\\
\sum_lG^l_mG^l_{m'}(\hat G \cdot \hat k)^2={3\over 8}|G|^2(P^T_{mm'}+3P^L_{mm'}),
\label{eq:B.12}
\end{eqnarray}
we obtain
\begin{eqnarray}
N^R_{ij}(k)&&=N_{ij}(k)+g^2\tilde a^4|G|^2T{3\pi\over 2}
\biggl({1\over\mu^2k^2}P^T_{ij}\bigl[I_0(\psi)+{1\over 2}I_1(\psi)\bigr]\nonumber\\
&&\phantom{N_{ij}(k)+g^2\tilde a^4|G|^2T{3\pi\over 2} }+
{1\over(2\mu+\lambda)^2k^2}P^L_{ij}\bigl[I_0(\psi)-{1\over 2}I_1(\psi)\bigr]
\biggr)
{1\over 4\tau}\biggl[\biggl({L\over\tilde a}\biggr)^{4\tau}-1\biggr].
\label{eq:B.13}
\end{eqnarray}
Upon projecting out of Eq.~(\ref{eq:B.13}) the longitudinal and 
transverse components, and recalling
that $|G|^2T=8\pi\mu{2\mu+\lambda\over 3\mu+\lambda}+O(\tau)$, we finally
arrive at the flow equations
\begin{eqnarray}
L{dA\over dL}&&=g^2{\tilde a}^4\biggl({L\over\tilde a}\biggr)^{4\tau} 
12\pi^2\mu {2\mu+\lambda\over 3\mu+\lambda}
\bigl[I_0(\psi)-{1\over 2}I_1(\psi)\bigr]\nonumber\\
L{dB\over dL}&&=g^2{\tilde a}^4\biggl({L\over\tilde a}\biggr)^{4\tau} 
12\pi^2\mu {2\mu+\lambda\over 3\mu+\lambda}
\bigl[I_0(\psi)+{1\over 2}I_1(\psi)\bigr].
\label{eq:B.14}
\end{eqnarray}

\section*{C} 
This Appendix details the calculation of 
the dynamical exponent $z$.
Consider the Langevin equation for the free hamiltonian $H_0$ of 
Eq.~(\ref{eq:10})
\begin{equation}
\gamma\partial_t\vec u(\vec r,t)=-{\delta H_0\over\delta\vec u(\vec r,t)}
+\vec\zeta(\vec r,t),
\label{eq:C.1}
\end{equation}
where $\zeta$ is a white noise defined by the correlation
\begin{equation}
<\zeta_i(\vec r,t)\zeta_j(\vec r',t')>=2D\delta_{ij}
\delta(\vec r-\vec r')\delta(t-t').
\label{eq:C.2}
\end{equation} 
The fluctuation-dissipation theorem insures that thermal equilibrium at
temperature $T$ is established for $D=T\gamma$.

Equation~(\ref{eq:C.1}) is easily solved.
The momentum and frequency dependent transverse and longitudinal 
correlation functions read
\begin{eqnarray}
S^{0T}_{ij}(\vec k,\omega)\equiv<u_i(\vec k,\omega)u_j(-\vec k,-\omega)>^T&&=
{2D\gamma^{-2}\over\omega^2+\gamma^{-2}\mu^2k^4}P^T_{ij}\nonumber\\
S^{0L}_{ij}(\vec k,\omega)\equiv<u_i(\vec k,\omega)u_j(-\vec k,-\omega)>^L&&=
{2D\gamma^{-2}\over\omega^2+\gamma^{-2}(2\mu+\lambda)^2k^4}P^L_{ij}.
\label{eq:C.3}
\end{eqnarray}

The presence of an external pinning potential turns Eq.~(\ref{eq:C.1}) 
into Eq.~(\ref{eq:27}) by introducing on the right hand side the additional term 
\begin{equation}
-{\delta U_{pin}\over\delta\vec u(\vec r,t)}=
{2V(\vec r)\over|\vec a\times\vec b|}\sum_{l=1}^3\vec G^l
\sin\biggl(\vec G^l\cdot\bigl(\vec r-\vec u(\vec r,t)\bigr)\biggr),
\label{eq:C.4}
\end{equation}
which can be regarded as a perturbation. 
First order perturbation theory can thus be employed to compute the
correlation functions. The perturbation series is found again to diverge 
logarithmically below $T_c$. This divergence causes the dynamical 
exponent to deviate from its mean field value, $z=2$ \cite{[23]}. To quantify this
deviation, consider the transverse correlation function 
\begin{eqnarray}
<u_i(\vec k,\omega)u_j(-\vec k,-\omega)>^T&&=
{1\over\omega^2+\gamma^{-2}\mu^2k^4}\biggl(2D\gamma^{-2}P^T_{ij}\nonumber\\
&&\phantom{{1\over\omega^2+\gamma^{-2}\mu^2k^4} }
+P^T_{im}P^T_{in}
\overline{<\hat F_m(\vec k,\omega)\hat F_n(-\vec k,-\omega)>}\biggr)+\dots,
\label{eq:C.5}
\end{eqnarray}
where the overline indicates averaging over the external pinning disorder
and we have defined 
\begin{equation}
\hat F_i(\vec k,\omega)=\int\!\! d^2r\int\!\! dte^{i\omega t}
e^{-i\vec k\cdot \vec r}{\delta U_{pin}\over\delta u_i(\vec r,t)}.
\label{eq:C.6}
\end{equation}
Upon carrying out the disorder average, we obtain
\begin{equation}
\overline{<\!\hat F_i(\vec k,\omega)\hat F_j(-\vec k,-\omega)\!>}=
\int\!\!dt e^{i\omega t}2g\!\sum_{l=1}^3G^l_iG^l_j
<\!\cos\vec G^l\cdot\bigl(\vec u(\vec r,t)-\vec u(\vec r,0)\bigr)\!>.
\label{eq:C.7}
\end{equation}
Noting that 
\begin{eqnarray}
<\!\cos\vec G^l\!\cdot\!\bigl(\vec u(0,t)\!-\!\vec u(0,0)\bigr)\!>&&=
\biggl({\tilde a\over L}\biggr)^{\!2-2\tau}\!\!\!\!
\exp\biggl(\!\sum_{ij}\!\!\int\!\!{d^2k\over 4\pi^2}G^l_i\bigl(
S^{0T}_{ij}(\vec k,t)+S^{0L}_{ij}(\vec k,t)\bigr)G^l_j\!
\biggr)\nonumber\\
&&=\biggl({\tilde a\over L}\biggr)^{2-2\tau}\!\exp\biggl[{|G|^2T\over 8}
\int_0^\infty\!\!dkk
\biggl({e^{-{\mu\over \gamma} k^2|t|}\over\mu k^2}+
{e^{-{2\mu+\lambda\over \gamma}k^2|t|}\over
(2\mu+\lambda)k^2}\biggr)\biggr]\nonumber\\
&&=\biggl({\tilde a\over L}\biggr)^{2-2\tau}
\biggl({\mu\over 2\mu+\lambda}\biggr)^{(1-\tau)\mu\over 3\mu+\lambda}
\biggl({L^2\over 2\sqrt{c}\gamma^{-1}\mu |t|}\biggr)^{1-\tau},
\label{eq:C.8}
\end{eqnarray}
and setting $\omega=0$, we arrive at
\begin{equation}
\overline{<\!\hat F_i(\vec k,\omega)\hat F_j(-\vec k,-\omega)\!>}=
2g{\tilde a}^2{3\over 2\mu}|G|^2\delta_{ij}
\biggl({\mu\over 2\mu+\lambda}\biggr){\mu\over 3\mu+\lambda}
{2\gamma\over\sqrt{c}}\log\biggl({L\over a}\biggr)+O(g\tau).
\label{eq:C.9}
\end{equation}
Thus, we can write
\begin{equation}
L{dD\over dL}=24\pi ga^2\biggl({L\over a}\biggr)^{2\tau}\sqrt{c}
{2\mu+\lambda\over 3\mu+\lambda}
\biggl({\mu\over 2\mu+\lambda}\biggr)^{\mu\over 3\mu+\lambda}\equiv \zeta(g).
\label{eq:C.10}
\end{equation}
The dynamical exponent is determined by the value of $\zeta(g)$ at the
fixed point
\begin{equation}
z=2+\zeta(g^*),
\end{equation}
leading to Eq.~(\ref{eq:28}).

\section*{D}  
The Ornstein-Zernicke theory of hexatics is derived in this
Appendix. We start from the partition function of the crystal
\begin{equation}
{\cal Z}=\int\!\! {\cal D}[u_1] {\cal D}[u_2]\exp(-F/T),
\label{eq:D.1}
\end{equation}
where 
\begin{equation}
F={1\over 2}\int d^2r (\lambda u_{ii}^2 +2\mu u_{ij}^2).
\label{eq:D.2}
\end{equation}
In the presence of dislocations, the displacement field $\vec u$ is
multivalued; if an arbitrary loop $\cal L$ encloses dislocations with total
Burger's vector $\vec b$, then
\begin{equation}
\oint_{\cal L}\partial_i u_j dx_i = b_j.
\label{eq:D.3}
\end{equation}
It is convenient to express Eq.~(\ref{eq:D.3}) in local form, 
\begin{equation}
\epsilon_{ik}\epsilon_{jl}\partial_k\partial_l u_{ij}(\vec r)
=\sum_\alpha\vec b^\alpha\times\vec\nabla\delta(\vec r-\vec r_\alpha)
\equiv S(\vec r).
\label{eq:D.4}
\end{equation}
Here, we have introduced a source term, $S(\vec r)$, related to the local
density of Burger's vectors; it can be easily generalized to include other
types of defects, such as disclinations, vacancies, or interstitials 
\cite{[24]}. 

The partition function can now be evaluated as an unresticted integral over
the three independent components of the symmetric strain tensor $u_{ij}$,
provided that the constraint (\ref{eq:D.4}) is enforced. This is easily accomplished
with the aid of an auxiliary field $\psi$:
\begin{equation}
{\cal Z}=\int\!\! {\cal D}[u_{11}] {\cal D}[u_{12}]{\cal D}[u_{22}]
{\cal D}[\psi] \exp(-F/T) 
\exp\biggl(i\int d^2r \psi (r)\bigl(
\epsilon_{ik}\epsilon_{jl}\partial_k\partial_l u_{ij}(\vec r)-S(\vec r)
\bigr)\biggr).
\label{eq:D.5}
\end{equation}

Next we write, as can be most generally done for any symmetric tensor,
\begin{equation}
u_{ij}(\vec r)={1\over 2}(\partial_i v_j+\partial_j v_i)+P^T_{ij}h,
\label{eq:D.6}
\end{equation}
where $v(\vec r)$ is a single-valued vector (describing, e.g., 
the small thermal displacements of the atoms from the sites of a
randomly distorted lattice), and $h(\vec r)$ a scalar field, related to the
defect density $S(\vec r)$, in terms of 
which the constraint enforced by the auxiliary field $\psi$ becomes
\begin{equation}
\nabla^2h(\vec r)=S(\vec r).
\label{eq:D.7}
\end{equation}

The partition function becomes a functional integral over $\vec v,\ h$, and
$\psi$, and the latter fields can be integrated out easily to obtain the
free energy functional
\begin{eqnarray}
{\cal F}&&={1\over 2}\int {d^2k\over (2\pi)^2} 
v_i(-\vec k)k^2\biggl(\mu P^T_{ij}+
(2\mu+\lambda)P^L_{ij}\biggr)v_j(\vec k)\nonumber\\
&&+\lambda\int d^2r(\vec\nabla \cdot\vec v)\nabla^{-2}S(\vec r)+
{1\over 2}(2\mu+\lambda)\int d^2r\bigl(\nabla^{-2}S(\vec r)\bigr)^2.
\label{eq:D.8}
\end{eqnarray}
This is a rather general expression of the two dimensional elastic free 
energy. It allows us to study the effect of both quenched and 
annealed defects.

First, we consider the effect of annealed dislocations.
It is instructive to recast Eq.~(\ref{eq:D.8}) in terms of density fluctuation
and bond angle variables, $\delta\rho$ and $\theta$ respectively, which
are related to the divergence and curl of the displacement vector:
$\delta\rho=\vec \nabla\cdot\vec u$, $\theta=\vec \nabla\times\vec u/2$.
Because the displacement vector decomposes naturally into single-valued
and singular parts, 
\begin{equation}
\vec u=\vec v+\vec u_s,
\end{equation}
so do the density fluctuation and bond angle variables:
\begin{equation}
\theta=\theta_0+\theta_s;\ \ \delta\rho=\delta\rho_0+\delta\rho_s.
\end{equation}
The defect source term, $S(\vec r)$, can be expressed in terms of $\rho_s$ by
noting first that, for a distribution of dislocations 
with Burger vector density $\vec b(\vec r)=\sum_\alpha\vec
b^\alpha\delta(\vec r-\vec r_\alpha)$, 
$S(\vec r)=-\vec \nabla\times \vec b(\vec r)$ from Eq.~(\ref{eq:D.4}), and hence, 
\begin{equation}
\nabla^{-2}S(\vec r)={1\over 1-\sigma}\vec \nabla\cdot\vec u_s=
{\delta\rho_s\over 1-\sigma},
\label{eq:D.9}
\end{equation}
where $\sigma=\lambda/(2\mu+\lambda)$ is the Poisson ratio.

The partition function of annealed dislocations \cite{[8]} involves averaging 
over thermally excited Burger's vectors with weight proportional to
\begin{equation}
\exp\biggl(-E_c\int d^2r\vec b(\vec r)\cdot \vec b(\vec r)\biggr)=
\exp\biggl(-E_c\int d^2r\bigl(4(\nabla\theta_s)^2
+{1\over (1-\sigma)^2}(\nabla\delta\rho_s)^2\bigr)\biggr).
\label{eq:D.10}
\end{equation}
Thus, from Eqs.~(\ref{eq:D.8}-\ref{eq:D.10}), it follows that 
\begin{equation}
{\cal Z}_{annealed}=\int\!\! {\cal D}\delta\rho_0{\cal D}\delta\rho_s
{\cal D}\theta_0{\cal D}\theta_s
\exp\biggl(-{{\cal F}_{annealed}\over T}\biggr)
\label{eq:D.11}
\end{equation}
with
\begin{eqnarray}
{\cal F}_{annealed}&&={1\over 2}\int d^2r \biggl(4\mu \theta_0^2+
(2\mu+\lambda)(\delta\rho_0)^2\biggr)
+{\lambda\over 1-\sigma}\int d^2r(\delta\rho_0)(\delta\rho_s)\nonumber\\
&&+{2\mu+\lambda\over 2(1-\sigma)^2}\int d^2r(\delta\rho_s)^2+
E_c\int d^2r\biggl(4(\nabla\theta_s)^2
+{1\over (1-\sigma)^2}(\nabla\delta\rho_s)^2\biggr).
\label{eq:D.12}
\end{eqnarray}

Consider now the linear response of the system to a probe coupled to the bond
angle. One finds easily that
\begin{equation}
{1\over \mu(q)}={1\over \mu}+{1\over E_cq^2},
\end{equation}
showing that the shear modulus $\mu$ vanishes at long wave lengths for finite
$E_c$ (i.e., above the melting temperature). Similar behavior for $\mu(q)$
was found by Marchetti and Nelson in a dislocation loop model of the melted
Abrikosov flux lattice\cite{[25]}. In contrast, the coupling between 
$\delta\rho_0$ and $\delta\rho_s$ prevents the bulk modulus from vanishing at
long wave lengths.

\begin{figure}
\caption{
(a): Directed lines in 1+1 dimensions subjected to random point pinning 
due to a disordered substrate.
(b): Tethered network of particles in 2+0 dimensions subjected to random 
point pinning due to a disordered substrate.}
\label{fig1}
\end{figure}
\begin{figure}
\caption{
(a): Annealed dislocation disorder embedded in an otherwise six-fold
coordinated membrane. Heavy lines join the 5- and 7-coordinated sites at the
cores of the dislocations. The disordered
substrate potential is not shown.}
(b): Random substitutional disorder in a polymerized membrane which
preserves the sixfold coordination of a perfect lattice. The disordered
substrate potential which acts on this lattice is not shown.
\label{fig2}
\end{figure}


\begin{references}
\bibitem{[thanks]} We are indebted to T. Hwa for discussions on this point.
\end{references}
\end{document}